\documentclass[reprint, prb, aps, amsmath, showpacs, superscriptaddress]{revtex4-2}
\usepackage{graphicx}
\usepackage{epsfig}
\usepackage[cp1251]{inputenc}
\usepackage[english]{babel}
\usepackage{amsmath}
\usepackage{amssymb}
\usepackage{amstext}
\usepackage{color}
\usepackage{float}
\usepackage{epstopdf}
\usepackage{bm}
\begin{document}

\title{Rotating bi-electron in two-dimensional systems with mexican-hat single-electron energy dispersion}

\author{V. A. Kochelap}
\email[E-mail:]{ kochelap@ukr.net}
\affiliation{Department of Theoretical Physics, Institute of Semiconductor Physics,National Academy of Sciences of Ukraine, Pr. Nauki 41, Kiev 03028,
Ukraine}

\begin{abstract}

A number of novel two-dimensional materials and nanostructures demonstrate
complex single-electron energy dispersion, which is called the {\it mexican-hat}
dispersion. In this paper, we analyze interaction of a pair of electrons
with such an energy dispersion. We show that relative motion
of  the electron pair is of a very peculiar character.
For example, the real space trajectories corresponding to electron-electron
scattering can have three reversal points,
reversal points at non-zero radial momentum and other unusual features.
Despite the repulsive Coulomb interaction, two electrons can be coupled forming
a composite quasi-particle - the {\it bi-electron}.
The bi-electron corresponds to excited states of the two-electron system.
Because the bi-electron coupled states exist in continuum of extended (free)
states of the electron pair, these states are quasi-resonant
and have finite times of life.  We found that rotating bi-electron is
a long-living composite quasi-particle.
The rotating bi-electrons can be in motion. For slowly moving bi-electrons, we determined
the kinetic energy and the effective mass.  Due to strongly nonparabolic
energy dispersion, the translational motion of the bi-electron is coupled to its
internal motion. This results in effective masses dependent on quantum states
of the bi-electron. In the paper, properties of the bi-electron are illustrated for
the example of bigraphene in a transverse electric field. We suggest that investigation of rotating bi-electrons at the mexican-hat single-electron
energy dispersion may bring new interesting effects in low-dimensional
and low-temperature physics.
\end{abstract}

\maketitle
\section{Introduction}\label{Introduction}

\begin{figure*}
\includegraphics{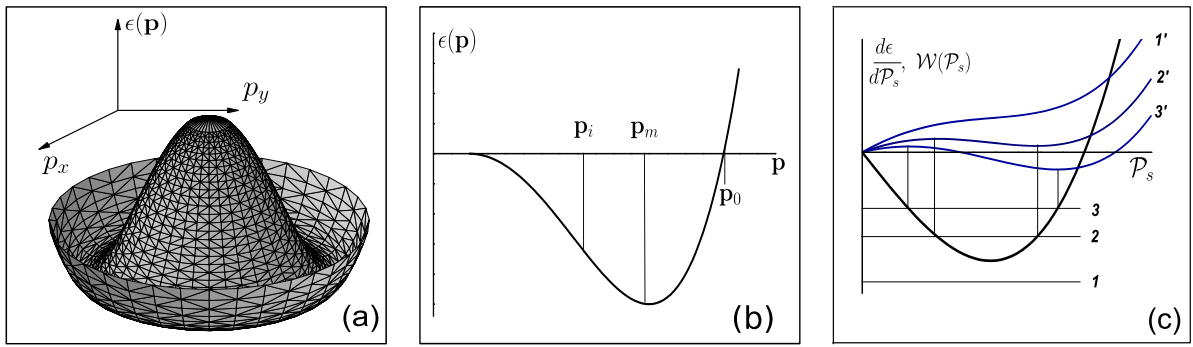}
\caption{{ {\bf a}: Sketch of the mexican-hat energy dispersion of a two-dimensional quasi-particle.
{\bf b}: The energy $\epsilon$ {\it vs} the momentum magnitude $p$.
{\bf c}: Illustrations to solutions of Eq.~(\ref{s-p-2}) -  the l.h.s., $d \epsilon/d {\cal P}_s$,
(the full curve) and the r.h.s., $\alpha/p_{\phi}$, (thin lines) at different $p_{\phi}$;
1 -  $p_{\phi,1} < p_{\phi,c}$;
2 - $p_{\phi,2} > p_{\phi,c}$; 3 - $p_{\phi,3}>p_{\phi,2}$. Curves 1', 2', 3' represent
the total energy, $E ({\cal P}_s)$, defined by Eq.~(\ref{r-e}) for
 $p_{\phi,1}$, $p_{\phi,2}$, $p_{\phi,3}$, respectively. For $p_{\phi} = p_{\phi,1}$, the line 1 does not cross the
$d \epsilon/d {\cal P}_s$ dependence and there are no singular points.
For $p_{\phi,2}$, $ p_{\phi,3}$, intersections of the lines 2, 3 and the
$d \epsilon/d {\cal P}_s$ dependence define the singular points.
Thin vertical lines show matching these singular points  and
extrema of the $E ({\cal P}_s)$-dependencies.}}
\label{fig-1}
\end{figure*}

It is well known that the crystalline potential affects motion of band electrons (holes)
strongly modifying their kinetic energy $\epsilon ({\bf p})$, where $\bf p$ is the electron
(hole) momentum. Typically, near an extremum, $\epsilon ({\bf p})$ is of a parabolic
dependence with an effective (generally anisotropic) mass. At larger $\bf p$ the
nonparabolicity is essential and a portion of the $\epsilon ({\bf p})$ dependence with
a negative effective mass ($d^2 \epsilon ({\bf p})/d p^2 <0$) can exist.
Graphene - atomically thin layer of the carbon atoms - provides the example of
striking kinetic energy  modification of both electrons and holes, when
the $\epsilon ({\bf p})$-dependencies have the linear quasi-relativistic behavior.
The bilayer graphene - bigraphene - represents even more complex behavior of low energy
$\epsilon ({\bf p})$-dependencies, which additionally can be controlled by external fields.
Recently, a few novel materials and nanostructures have been fabricated
for which, in the lower energy bands, the two-dimensional electrons are characterized
by complex $\epsilon ({\bf p})$-dependencies that can be called the "mexican-hat"
(MH) energy dispersion. For the MH energy dispersion, a local {\em maximum} occurs say
at, ${\bf p} =0$ and a portion with the negative effective mass exists near this
maximum.  A {\em minimum} value of  $\epsilon ({\bf p})$ is reached at a circle $|{\bf p}| =  p_m$,
at larger $\bf p$  the function $\epsilon ({\bf p})$  increases.
The MH energy dispersion is sketched in Fig.~\ref{fig-1}~(a).
Examples of two-dimensional materials and nanostructures with this type of the energy dispersion
include: bigraphene in a transverse electric field (both, electron and hole
bands)~[\onlinecite{bigraphene-1}], [\onlinecite{bigraphene-2}],
hole bands in few-layers III-VI materials, such as GaSe, GaS, InSe, InS and
Bi$_2$Te$_3$, Bi$_2$Se$_3$~[\onlinecite{III-VI-compounds-1,III-VI-compounds-2,III-VI-compounds-3,III-VI-compounds-4}].
Also, the MH energy dispersion is characteristic for HgTe-HgCdTe quantum wells
(the upper hole  band)~\cite{HgCdTe-2,HgCdTe-3},
InAs/GaAs double quantum well structures~[\onlinecite{InAs-GaSb}],
the strained quantum well structures fabricated by III-V compound (the upper hole band),
including strained GaAs/AlGaAs and GaN/AlGaN
structures, etc. These examples indicate that the single-electron energy dispersion of
the MH type is quite general phenomenon, especially for two-dimensional systems.

For materials with the MH energy dispersion
relative motion of  the electron pair is very specific. Particularly,
a repulsive interaction potential can lead to electron pairing, i.e.
to formation of a composite quasi-particle - {\em a bi-electron}.

The term "bi-electron" is known in solid-state physics. It was introduced for
the case of coupling of two electrons originated from different energy bands,
one of them from the bottom of the lower conduction band
and the other  from the top of an upper band with negative curvature. So that,
the reduced effective mass of the pair can be negative, which can give rise
to electron  pairing in spite of
Coulomb repulsion.  In particular, the bi-electron model was applied to
explain the inverse hydrogen-like series of optical lines observed in
layered $BiI_3$ crystals~[\onlinecite{Gross}]. Another example of formation of
bi-electrons near the saddle points of the two-particle energy
dispersion in strong magnetic fields was analyzed in paper~[\onlinecite{Rashba}]
(quasi-one-dimensional bi-electrons). In both mentioned examples, coupled
electrons were originated from different energy bands, the models of the energy
dispersion were restricted to parabolic dependencies. Note, in composite structures
- semiconductor/metal - bi-electron can be formed due to both image forces and
spin-orbit interaction~[\onlinecite{Chaplik}].

Recently, investigations of electron-electron interaction in graphene-based
two-dimensional systems~[\onlinecite{Sabio}]
 sparked the interest to the bi-electron problem.
Indeed, in paper~[\onlinecite{Lee}] quasi-localized states of two electrons were found
to be possible. Then, the study presented in paper~[\onlinecite{Shytov-1}]  showed
that two-electron states can exist if  single-electron energy dispersion deviates
from the linear Dirac-like spectrum.
In the cited paper, two-particle states were found for double-layer graphene
structures  in the model with additional quadratic momentum term with
a negative effective mass. This additional term was derived by taking into account
the hopping of the electrons between next-nearest-neighbor atoms. Absolute
value of the effective mass introduced by such a way was found about five time larger
than the free electron mass. Being applied to single layer or bilayer graphene
structures,  such an approach led to unreasonably large coupling energy ($\sim 1~eV$).
However, the same model applied to graphene layers separated by boron nitride
with the interlayer Coulomb potential led to the coupling energies of order of
tens $meV$. Further studies of this subject were focused on electron-electron interaction
in topological insulators~[\onlinecite{Sablikov}], electron pairing was analyzed in the
four-band model assuming a step-like repulsive potential.

In this paper, we revisite problems of the interaction of two electrons and
bi-electron states formation in two-dimensional systems.
We  assume that coupling energies of pairing electrons are much smaller, than
energy separation of considered single-electron band of the MH type from other electron
bands.  We show that, despite the repulsive Coulomb interaction, the two electrons
with the single-electron MH dispersion can be coupled forming excited states of the
two-electron system.
We found that the rotating bi-electrons are of long-living quasi-particles.
The rotating bi-electrons can be in translation motion. For slowly moving bi-electron
we determined the kinetic energy and the effective mass. The presented model,
which exploits the single-electron energy band,  facilitates the analysis of
other important properties of two-dimensional bi-electrons.

The model under consideration is formulated as follows.
We consider two-dimensional electron system, when the electron momentum is
${\bf p}=\{p_x,p_y\}$.
We assume that the $\epsilon ({\bf p})$-dependence is of the MH type, as illustrated in Fig.~\ref{fig-1} (a).
This energy is an isotropic function in the $\{p_x,p_y\}$-plane. For such a type of the
$\epsilon ( p)$-dependence, one can introduce a few characteristic parameters:
the inflection point, $p_i$ ($d^2 \epsilon /d p^2 =0$ at $p=p_i$); the momentum $p_m$,
corresponding  to the energy minimum ($d \epsilon / d p =0$ at $p=p_m$) and
the momentum  $p_0 $, corresponding to zero energy ($\epsilon (p_0) =0 $ at $p_0  \neq 0$),
as illustrated in Figs.~\ref{fig-1} (b).

For a pair of the interacting electrons, say 1 and 2, the Hamiltonian  is
\begin{equation} \label{H1}
 H = \epsilon ( {\bf  p}_1) + \epsilon ({\bf p}_2) + U(|{\bf r}|) \,,
\end{equation}
where $U (|{\bf r}|) > 0$ is the potential of the electron-electron interaction, and
${\bf r} = {\bf r}_1 -{\bf r}_2$ is the relative distance between
the electrons. Introducing the total and relative momenta for the two-electron system,
$\bf{P}=\bf{p}_1+\bf{p}_2\,$ and $\bf{p} = ({\bf p}_1- {\bf p}_2) / 2$, respectively,
we rewrite Eq.~(\ref{H1}) as $ H = \epsilon \left( {\bf P}/{2} + {\bf  p}\right)
 + \epsilon \left({\bf P}/{2}- {\bf p}\right) + U(|{\bf r}|) \,.$
In the absence of external lateral fields in the $\{x,y\}$-plane, the total momentum
is conserved, i.e., ${\bf P} =constant$.
Let the center-of-mass of the pair be motionless, i.e., ${\bf P}=0$. Then, the Hamiltonian
corresponding to relative motion of the two electrons takes the form:
\begin{equation} \label{H3}
H_0 =2  \epsilon ( {\bf  p})  + U(|{\bf r}|) \,.
\end{equation}
This is the basic equation for quantitative analysis given in the next Sections.

Here, we may use Eq.~(\ref{H3}) to explain qualitatively two-electron pairing effect at
the MH energy dispersion.   Indeed, near the point ${\bf p}=0$ one can simplify $H_0$ to
the form $H_0 \approx - {\bf p}^2/2 |M| +U$, with $M =1/( d^2 \epsilon ({\bf p})/d p^2|_{p=0}) $
being the negative "reduced" effective mass. The auxiliary Hamiltonian,
$H_0' = - H_0 = {\bf p}^2/2 |M| - U$, describes attractive particles, that can have coupled
states with 'energies' $E'< 0$.  Comparing $H$ and $H'$, we can expect that for the
Hamiltonian of Eq.~(\ref{H3}), coupled states may exist at energies $E >0$.
Since the pairing effect is due to the negative effective mass,
 radii of coupled states (in real space) have to be large enough to provide
(in the momentum representation) the main contribution from small relative momenta ${\bf p}$,
where the negative effective mass occurs.
Simultaneously, at energies $E>0$ there exist also uncoupled states
of the electron pair, corresponding to electron-electron scattering. For uncoupled states,
the main contribution comes from finite momenta ${\bf p}$ (in the momentum representation).
In the semiclassical picture, coupled and uncoupled motion of the two
electrons at a given  energy are independent: they correspond to
different initial conditions. While, in the quantum picture,
there exists a tunneling between states of the same energy.
Therefore, sought-for states have to be {\em quasi-coupled} and be characterized by
a finite decay time. Note, under rotation of the electron pair a finite centrifugal potential
gives rise to an increase of the radius of a coupled state, that, in turn, rises its decay time.
Concluding this qualitative consideration, one can expect that
in systems with the MH type electron energy it is possible the formation of
 composite quasi-particles - bi-electrons. Bi-electron states are {\it excited} and metastable
 states  of the two-electron system.  Rotating bi-electrons should be long-living quasi-particles.

The rest of the paper is organized as follows. In Section \ref{semiclassical},
we present semiclassical analysis of the problem and give a classification of
possible patterns of two-electron motion, and illustrate the
results by a few particular models of the MH energy dispersion. In Section
\ref{Quantum analysis}, we develop a quantum approach to the problem,
determine energies, wavefunctions, decay times, and spins of the bi-electrons.
Finally in this Section, we consider moving bi-electron.
In Section \ref{Discussion}, we discuss the obtained results and present
numerical estimates with focus to the particular example - the bi-layer graphene
subjected to a transverse electric field.
A short summary of the overall results is presented in Section \ref{Summary}.

\section{Semiclassical consideration} \label{semiclassical}

\subsection{Equations for relative motion of two electrons}
We start with the semiclassical analysis. For such a case, the Hamiltonian
of Eq.~(\ref{H3}) is a function of two variables, absolute values of the two-dimensional
vectors $\bf p$ and $\bf r$: $H_0=H_0 (p,r)$.
It is convenient to use the polar coordinates, $\{ r, \phi \}$,
instead of the orthogonal coordinates, $\{ x, y \}$, i.e., $x = r cos \,\phi \,,\, y = r sin\, \phi $.
Then, instead of $ p_x$ and $ p_y$ we introduce $p_r =p_x cos\, \phi + p_y sin \, \phi $ and
$p_{\phi}= r (p_y cos \, \phi - p_x sin \, \phi)$ with $p^2 = p_r^2 + p_{\phi}^2/r^2 $.
Obviously, $p_{\phi}$ is the angular momentum of the pair of the electrons.
In the new variables $\{r, \phi \}$ and $\{p_r, p_{\phi} \}$, Eq.~(\ref{H3}) reads
\begin{equation} \label{H4}
H_0 (p_r,p_{\phi},r,\phi) = 2 \epsilon \left( \sqrt{p_r^2 + p_{\phi}^2/r^2} \right) + U(r)\,.
\end{equation}
The corresponding equations of motion are:
 \begin{equation} \label{r-t}
\frac{d r}{d t} = \frac{\partial H_0}{\partial p_r}= \frac{2 p_r}{ \cal P} \frac{d \epsilon}{d {\cal P}}
\equiv {\cal R}_r (p_r, r |p_{\phi})\,,
\end{equation}
\begin{equation} \label{pr-t}
\frac{d p_r}{d t} = - \frac{\partial H_0}{\partial r}=  \frac{2 p_{\phi}^2}{r^3 \cal P} \frac{d \epsilon}{d {\cal P}}
- \frac{d U}{d r} \equiv {\cal R}_p (p_r, r |p_{\phi})\,,
\end{equation}
\begin{equation} \label{phi-t}
\frac{d \phi}{d t} = \frac{\partial H_0}{\partial p_{\phi}}= \frac{2 p_{\phi}}{r^2 \cal P}
 \frac{d \epsilon}{d {\cal P}} \,,
\end{equation}
\begin{equation} \label{pphi-t}
\frac{d p_{\phi}}{d t} = - \frac{\partial H_0}{\partial \phi}=0\,.
\end{equation}
In these equations, $t$ is the time, $\frac{d \epsilon}{d {\cal P}}$ stands for the
derivative of the function $\epsilon(p)$ calculated at
$p={\cal P} \equiv \sqrt{p_r^2 + p_{\phi}^2/r^2}$.
The repulsive potential is supposed to be of the Coulomb type:
\begin{equation} \label{coulomb}
U(r) = \frac{\alpha}{r}\,,\,\,\alpha >0\,,
\end{equation}
where $\alpha$ depends on the dielectric environment, ${\cal R}_r$ and ${\cal R}_p$
are designations for the right hand sides of Eqs.~(\ref{r-t}) and (\ref{pr-t}),
respectively.

The system of Eqs.~(\ref{r-t})-(\ref{pphi-t}) has the following properties.
Eq.~(\ref{pphi-t})  implies $p_{\phi} = const$, which means conservation of
the angular momentum.
Eqs.~(\ref{r-t})-(\ref{pr-t}) do not depend on the angle $\phi$.
Thus, the radial motion of the pair described by Eqs.~(\ref{r-t}), (\ref{pr-t}) and its angular motion
described by Eqs.~(\ref{phi-t})-(\ref{pphi-t}) are decoupled.
If the radial variables, $r (t)$ and $p_r(t)$, are found, then $\cal P$ and
$d \epsilon/d {\cal P}$ are known functions of the time, $t$, and the angle
variable, $\phi (t)$, can be easily calculated by using Eq.~(\ref{phi-t}).

\begin{figure}
\includegraphics{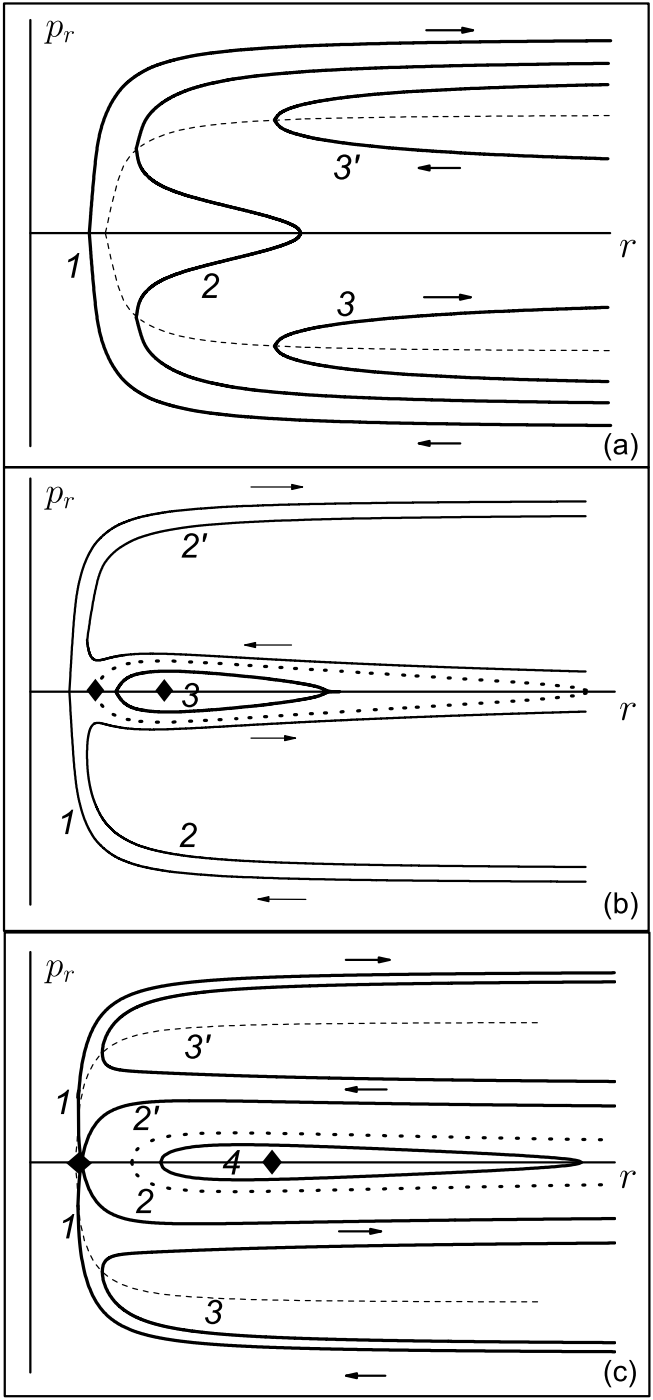}
\caption {{Phase portraits of Eq.~(\ref{phase-plane-1}) at different values of the
angular momentum, $p_{\phi}$. Solid lines are trajectories,
dashed lines (in (a) and (c)) correspond to equation
$\sqrt{p_r^2 + p_{\phi}^2/r^2 }= p_m$.
The arrows indicate direction of motion along the trajectories.
{\bf a}: The case $p_{\phi} < p_{\phi,c}$; curves 1, 2
are scattering trajectories of the type I and II,
respectively.
Trajectories are presented for different values of the total energy, $E$:
curve 1 for $E>E_{c1}$, the type I;
curve2 for $E_{c1} > E >0$, the type II;
curves 3, 3' for $E<0$, the type III.
{\bf b}: The case $p_{\phi}>p_{\phi,c}$ and $E^{s1}, E^{s2}>0$; two singular points are marked by
stars. Dotted line is the separatrix closed loop restricting a region of the phase plane with
closed trajectories.
Curves 1, 2, 2'  are trajectories of the scattering types I and III;
curve 3 is an example
of the closed trajectories with $E^{s1} > E > E^{s2} $.
{\bf c}:  The case $p_{\phi}>p_{\phi,c}$ and $E^{s1}>0, E^{s2} <0$;
curves 1, 2, 2', 3, 3' are
trajectories of the scattering types. The dotted line restricts a region
with closed trajectories.
Curve 4 is an example of  the closed trajectory for $E^{s1} > E >0$.
}}
\label{fig-2}
\end{figure}

\subsection{Phase-plane analysis}
We shall focus on the equations for the radial motion (\ref{r-t}), (\ref{pr-t}), which
compose {\it an autonomous system} of differential equations for the variables $r$
and $p_r$.
Such a system can be studied in details by using the so-called
phase-plane analysis (see, for example, Ref.~ [\onlinecite{Kevorkian}]).
The phase-plane analysis is based
on simple reduction of this system to a single differential equation of the first order:
\begin{equation} \label{phase-plane-1}
\frac{d p_r}{d r} = \frac{{\cal R}_p (p_r, r |p_{\phi})}{{\cal R}_r (p_r, r |p_{\phi})}\,,
\end{equation}
where the right-hand side (r.h.s) is parametrically dependent on $p_{\phi}$
~[\onlinecite{comment-2}].

At a given $p_{\phi}$, any solution of Eq.~(\ref{phase-plane-1}) can be presented as a certain
"trajectory" in the $\{r, p_r \}$-phase plane. Every trajectory corresponds to a certain
total energy, $E$. The energy conservation law gives the implicit equation of
these trajectories,
$H(p_r,r) = E$. A chosen point in the $\{r, p_r \}$-phase plane can be interpreted
as an initial condition attributed to a time moment $t'$. The trajectory which crosses
this point determines the dynamics of the electron pair at $t > t'$.
By using Eqs.~(\ref{r-t}), (\ref{pr-t}), it is easy to determine
the direction of the trajectories. The singular points of
Eq.~(\ref{phase-plane-1}), if any, correspond to a motion  of the pair with
time-independent $r$ and $p_r$
(a rotation in real space with a fixed angular velocity according to Eq.~(\ref{phi-t})).
The singular points can be either stable or unstable. Equating the numerator
and denominator in Eq.~(\ref{phase-plane-1}) to zero, we obtain two isoclinal lines: a trajectory $p_r (r)$
crosses the first isoclinal line always horizontally ($d p_r/d r = 0$), while the second
isoclinal line is always crossed vertically ($d p_r/d r \rightarrow \infty$). These properties of the
trajectories allow one to reconstruct readily the topology of the phase plane and to
study possible types of semiclassical relative motion of the pair of the electrons.

For the singular points  $(r^s,p_{r}^s)$, we obtain the equations
\begin{equation} \label{s-p-1}
p_{r}^s = 0 \,,
\end{equation}
\begin{equation}    \label{s-p-12}
\frac{2 p_{\phi}}{(r^s)^3} \frac{d \epsilon}{d {\cal P}} - \frac{d U}{d r} \big|_{r^s}=0\,\,\,
\mbox{at} \,\,{\cal P}= {{\cal P}^s}=\frac{p_{\phi}}{r^s}\,.
\end{equation}
For a repulsive potential, we have $d U/d r <0$, thus Eq.~(\ref{s-p-12})  may have
solutions only at $d \epsilon/d {\cal P} < 0$. In the case of the energy
dispersion shown in Fig.~\ref{fig-1} (b), this leads to the conditions: ${\cal P} ^s < p_m$ and $r^s > p_{\phi}/p_m$.
Then, using Eq.~(\ref{coulomb})  and the relationship between ${\cal P}^s$ and $r^s$
we obtain the following simple equation for ${\cal P}^s$:
\begin{equation} \label{s-p-2}
2 \frac{d \epsilon}{d {\cal P}}\big|_{{\cal P}^s} = - \frac{\alpha}{p_{\phi}} \,.
\end{equation}
Here only the left-hand side (l.h.s.) is varied with ${\cal P}^s$, the r.h.s.  is
negative and depends parametrically on the angular momentum $p_{\phi}$.
For the MH energy dispersion, the l.h.s. is negative at ${\cal P}^s < p_m$ and
reaches a minimum at the inflection point, ${\cal P}^s= p_i$.
Graphical solutions of Eq.~(\ref{s-p-2}) are illustrated in
Fig.~\ref{fig-1} (c).
These solutions arise at
\begin{equation} \label{p-phi-c}
p_{\phi} \geq p_{\phi,c} =  {{\alpha}} \bigg/ \left[ 2 \bigg|  \frac{d \epsilon}{d {\cal P}} \bigg|_{p_i} \right]\,.
\end{equation}

At $p_{\phi} < p_{\phi,c}$,  Eq.~(\ref{s-p-2}) has no solutions and, thus, there are
not singular points (see the illustration in Fig.~\ref{fig-1} (c).
The corresponding phase portrait of Eq.~(\ref{phase-plane-1}) is shown in
Fig.~\ref{fig-2} (a).
Trajectories starting and ending at $r \rightarrow \infty$ correspond to  processes of
electron-electron scattering and can be called as {\em scattering trajectories}.
Depending on the total energy, $E$, there are three types of the scattering trajectories.
The type I encloses the trajectories of the energy $E > E_{c1} $,
\begin{equation} \label{add-1}
E_{c1} = 2 \epsilon( p_m) + \frac{\alpha}{p_{\phi}} p_m~.
\end{equation}
Each of these trajectories is a continuous line, which starts at
$r \rightarrow \infty$ and $p_r \rightarrow  - p_r (E)$ (with $p_r(E)$
that satisfies the condition $2 \epsilon (p_r) = E$),
passes through lower and upper parts of the phase-plane, and finishes with positive
$p_r \rightarrow p_r (E)$ at $r \rightarrow \infty$.
The trajectory crosses the $r$-axe only once at
the coordinate $r=r_{E}$, which can be found from the equations:
\begin{equation} \label{r-e}
r_{E} = \frac{p_{\phi}}{{\cal P}_E}\,,\,\,\,\,E = 2 \epsilon({\cal P}_E) +
\frac{\alpha}{p_{\phi}} {\cal P}_E \equiv W({\cal P}_E)\,.
\end{equation}
For this type of the trajectories, $W ({\cal P}_E)$-dependence is illustrated in
Fig.~\ref{fig-1} (c) by the curve $1'$. The trajectories have a single {\em reversal point}
$(r_{E},\,0)$ with $r_{E} < p_{\phi}/p_m$. At the reversal point, the radial velocity changes
its sign. The type I of the phase-plane trajectories corresponds to usual processes of
elastic scattering in real space.

The type II encloses trajectories of the energy interval $0 <E < E_{c1} $. At large $r$ they behave similarly to the type I, however, they have  unusual feature at finite $r$.
Indeed, for these trajectories there exist {\em three reversal points}. For each of such trajectories one of the reversal point is on the $r$-axis and is determined by
Eqs.~(\ref{r-e}). Two other are on the line $p_r = \pm \sqrt{ p_m^2 - p_{\phi}^2/r^2 }$.
For a given energy $E$, the additional reversal points occur at
\begin{eqnarray} \label{r-e-add}
r_{r,E}^{ad} = \frac{\alpha}{E-2 \epsilon (p_m)}\,,  \nonumber \\
p_{r,E}^{ad} =\pm \sqrt{p_m^2 -\frac{p_{\phi}^2}{\alpha^2}\left[E- 2 \epsilon(p_m) \right]^2}\,.
\end{eqnarray}
From Eq.~(\ref{phi-t}), it follows that at the additional reversal points, the angular
velocity, $d \phi/ d t$, changes its sign.

The trajectories with $E< 0$ never cross the $r$-axis, they can be attributed to the type III.
For this case, there are two solutions of the equation $2 \epsilon (p) =E$, which we denote
$p^{m} (E)$ and $p^{M} (E)$ with $p^m < p_m < p^M$. This implies the existence of two isolated trajectories for a given negative $E$. One trajectory starts at infinity ($r \approx \infty$)
with $p_r= - p^M (E)$. It has a reversal point given by  Eq.~(\ref{r-e-add}) with "-" sign in the second equation, remains in the lower part of the $\{r,\,p_r \}$-plane,  and finishes at infinity with $p_r= - p^m (E)$. Another one starts at infinity with $p_r = + p^m (E)$, and finishes at infinity with $p_r = + p^M (E)$. Corresponding reversal point is determined by
Eq.~(\ref{r-e-add}) with "+" sign. Despite the elastic character of the processes,
such trajectories describe collisions that give rise to {\em a change of the relative
momentum}, $p_r$, of the electron pair.

Now, we return to the case when the inequality (\ref{p-phi-c}) holds and
Eq.~(\ref{phase-plane-1}) has singular points.
The single such point,  $(r_s=p_{\phi}/ p_i, p_r=0)$
appears in the phase-plane at $p_{\phi} = p_{\phi,c}$.
When $p_{\phi} > p_{\phi,c}$,
Eq.~(\ref{s-p-12}) has two solutions ${\cal P}^{s1}$ and ${\cal P}^{s2}$ with
${\cal P}^{s1}< p_i < {\cal P}^{s2} < p_m$ (see illustration in
Fig.~\ref{fig-1} (c)).
Thus, there are two singular points $(r^{s1},0)$ and $(r^{s2},0)$, $r^{s1} > r^{s2} $.
Near a singular point $(r^s,0)$, the trajectories corresponding to different  energies $E$
can be found in the form:
\begin{equation}  \label{s-p-4}
\frac{1}{{\cal P}^s}\frac{d \epsilon}{{d \cal P}^s} \, p_r^2 + \frac{({\cal P}^s)^4}{p_{\phi}^2}
\frac{d^2 \epsilon}{d ({\cal P}^s)^2} (r-r^s)^2 = E-E^s\,,
\end{equation}
where $E^s$ is the total energy of relative motion of the electron pair
in the $s$-th singular point defined by the second equation from (\ref{r-e}):
$E^s= E({\cal P}_s)$.
For the $s1$-point with $\frac{d \epsilon}{{d \cal P}^s}<0$ and
$\frac{d^2 \epsilon}{d ( {\cal P}^{s })^2 }< 0$ (see Fig.~\ref{fig-1} (c)),  from Eq.~(\ref{s-p-4})
it follows that allowed energies are $E < E^s$ and the trajectories are closed curves.
That is, the $s1$-point is {\em the center}. While for the $s2$-point with
$\frac{d \epsilon}{{d \cal P}^s}<0$ and $\frac{d^2 \epsilon}{d ({\cal P}^s)^2} > 0$,
the trajectories are hyperboles. This $s2$-point is {\em a saddle}.
The appearance of the singular points leads to restructuring of the phase-plane.
There can exist  two cases of different phase-plane topologies.

For the first case, the phase-plane is presented in Fig.~\ref{fig-2} (b). For this case,
both singular points correspond to positive total energies, $E^{s1}, E^{s2}>0$,
defined by Eq.~(\ref{r-e}) at ${\cal P}_E = {\cal P}^{s1},\, {\cal P}^{s2}$, as illustrated by
the curve $2\,'$ in Fig.~\ref{fig-1} (c). In the phase-plane on
Fig.~\ref{fig-2} (b), two separatrices of the saddle form a closed loop
that restricts a finite region of the phase-plane where all the trajectories are closed.
For them the total energy $E$ is in the range $E^{s1} > E > E^{s2} >0$.
For a given energy from this range, the minimal and maximal coordinates, $r_m(E)$ and $r_M (E)$,
which can be reached on the closed trajectory, are to be found from Eqs.~(\ref{r-e}).
Note, for the same energy range there are trajectories of a scattering type. The latter
are well separated  from the closed ones, as illustrated in
Fig.~\ref{fig-2} (b) by curves
$1,\,3$. Outside of the discussed energy range, all the trajectories
are of the scattering types, as was found in the previous analysis.

Another case of the phase-plane topology is shown in Fig.~\ref{fig-2} (c). It occurs for
$E^{s1} >0 >E^{s2}$, as illustrated by the curve $3'$ in Fig.~\ref{fig-1} (c).
At other fixed parameters, this case corresponds to larger angular momenta.
Now the separatrices of the $s2$-saddle are extended up to infinity; they
do not form a closed loop. Instead, closed trajectories exist for the energy
interval $E^{s1} \geq E > 0$. For $E \rightarrow +0$ ,
these closed trajectories are extended to infinitely large $r$.
With increase in $p_{\phi}$, the singular points move toward larger $r$. At
$$
p_{\phi} \gg \frac{\alpha}{2 p_i \left| \frac{d^2 \epsilon}{d p^2} \right|_{p=0} }\,,
$$
 one can obtain $r^{s1} \approx 2 \left|{d^2 \epsilon}/{d p^2} \right|_{p=0}
{p_{\phi}^2}/{\alpha} \propto p_{\phi}^2$ and
$r^{s2} \approx p_{\phi}/p_0 \propto p_{\phi}$.
The region of the phase-plane, which contains  the closed trajectories, also is shifted
toward larger $r$. Under discussed strong inequality, the Hamiltonian of
Eq.~(\ref{H4}) is simplified to the form
\begin{equation} \label{Kepler-2}
H_0  = - \left|\frac{d^2 \epsilon}{d p^2} \right|_{p=0} \left(p^2_r +
\frac{p^2_{\phi}}{r^2} \right) + \frac{\alpha}{r}\,.
\end{equation}
Following the discussion presented in Introduction, one can introduce the auxiliary
Hamiltonian $H'_0= - H_0$, that can easily be reduced to that of the well known
Kepler problem.  Detailed analysis of this problem
can be found elsewhere~[\onlinecite{Landau-M}]. Particularly, from this analysis, it
follows that the closed trajectories exist for the energy interval
$0<E<E^{s1}=\alpha/4 p^2_{\phi} \left| {d^2 \epsilon}/{d p^2} \right|_{p=0}$.

\begin{figure}
\includegraphics{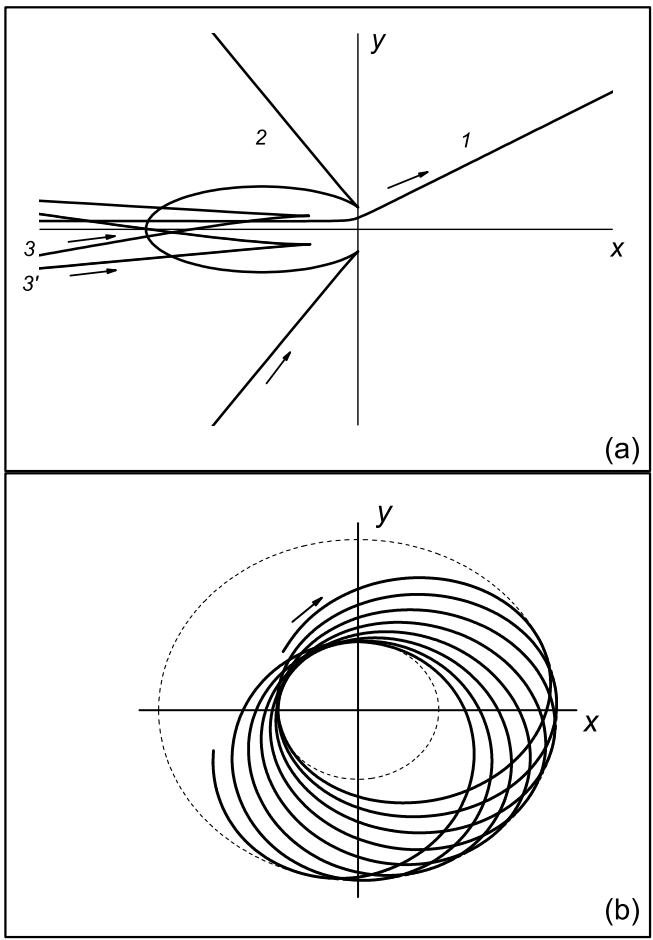}
\caption{{Trajectories in  real space.
{\bf a}: Scattering trajectories; curves $1,\,2$ and $3,\,3'$ - are scattering trajectories of the
types I, II, and III, respectively; they corresponds to the
phase-plane trajectories $1,\, 2$ and $ 3,\,3'$ shown in Fig.~\ref{fig-2} (a).
{\bf b}: A real-space trajectory illustrating bounded motion of the electron pair.
Dashed circles mark the minimal and maximal distances between the electrons.
}}
\label{fig-3}
\end{figure}

Now consider briefly relative motion of the pair of the electrons in {\em real  space}.
Trajectories in  the $\{x , y\}$-real space can be calculated, when solutions of
Eqs.~(\ref{r-t}), (\ref{pr-t}), for $r(t)$ and $\phi(t)$,
are found. Examples of the scattering trajectories are shown in
Fig.~\ref{fig-3} (a).
Among the presented curves, only trajectory $1$ has a standard form for the process
of scattering by a repulsive potential. Shapes of the others  are rather unusual
due to the complex energy dispersion of the MH type. For example, the curve $2$ represents
scattering trajectory with three reversal points, the curves $3,\,3'$ represent
 trajectories with single reversal points occurring at a finite radial momentum
   (matching these
real-space trajectories to those of the phase-plane is indicated in the caption).
Note, among trajectories belonging to the type II there are self-crossing trajectories
in real space (not shown in Fig.~\ref{fig-3} (a)).

The closed trajectories in the $\{r,p_r\}$-phase-plane correspond to relative
motion of two electrons, which occurs in a restricted region of real space.
At a given energy $E$, the real space trajectory lies in a ring bounded by the circles of
radii $r_m(E)$ and $r_M(E)$, both were defined above. Generally, these
trajectories are not closed, as illustrated in Fig.~\ref{fig-3} (b). At a given
angular momentum $p_{\phi}$,  the only closed real-space trajectory  is a
circulate orbit  with radius corresponding to
the $s1$ singular point: $r=r^{s1}$ at $E=E^{s1}$.  The electron pair moves round this
orbit with velocity $2  \left|d \epsilon/d p \right|_{p=p_{\phi} /r^{s1}}$.
The corresponding rotation frequency is
$\Omega^{s1} = 2/(r^{s1} \,   \left|d \epsilon/d p \right|_{p=p_{\phi}/r^{s1}})$.
For the motion with energy $E$ close to $E^{s1}$, the rotation occurs with small radial vibrations.
The frequency and the magnitude of these vibrations are equal to
\begin{equation} \label{small-vibr}
\Omega_r = \frac{2 {p_{\phi}}^{1/2}}{{r^{s1}}^{3/2}}
\sqrt{\left|\frac{d \epsilon}{d p}\, \frac{d^2\epsilon}{d p^2}\right|}\,,\,\,
r_M-r_m = 2 \frac{r_{s1}^2}{p_{\phi}} \sqrt{\frac{E^{s1}-E}{\left|d^2 \epsilon/d p^2\right|}} \,,
\end{equation}
where $E < E^{s1}$ and the derivatives are calculated at $p=p_{\phi}/r^{s1}$.
Obviously, spatially bounded relative motion of the two electrons means their coupling
despite the repulsive interaction.

Summarising, the general analysis showed that the two-electron coupling
in  real space arises with the onset of the singular points, $s1,\,s2$,
in the $\{r, \,p_r \} $ phase-plane of Eq~(\ref{phase-plane-1}).
Eqs~(\ref{s-p-1}), (\ref{s-p-12}) and (\ref{r-e}) for these points
facilitate the determination of the energies of $E^{s1},\,E^{s2}$, corresponding
to these points at a given angular momentum, $p_{\phi}$. The
two-electron coupling is realized for the following interval of
the total energy: $max\{0,\,E^{s2}\} < E < E^{s1}$.

\subsection{Numerical estimates} \label{N-estimates}

It is useful to conclude the semiclassical analysis with numerical  estimates.
We rescale the variables as follows:
\begin{eqnarray} \label{scaling}
{\bf k} = \frac{\bf p}{p_B},\,l_{\phi} = \frac{p_{\phi}}{p_{\phi,B}},\,
{\bm \rho}= \frac{\bf r}{r_B},\,
{\varepsilon} ({\bf k})= 2 \frac{\epsilon({\bf p})}{E_B},\,\\
{\cal H}_0 = \frac{H_0}{E_B}={\varepsilon}({\bf k}) + \frac{1}{\rho} \,,\nonumber
\end{eqnarray}
where we introduce the Bohr-like units:
\begin{eqnarray} \nonumber
p_B = \frac{M \alpha}{\hbar},\,\,r_B=\frac{\hbar}{p_B},\,\,
E_B=\frac{M \alpha^2}{\hbar^2},\,\,\\
M=\frac{ 1}{2 \left|d^2 \epsilon/d p^2 \right|_{p=0}},\,\,
p_{\phi,B} = p_B r_B=\hbar\,.\nonumber
\end{eqnarray}
Here $M$ is the reduced mass of the pair at $p \rightarrow 0$.
In dimensionless form, Eqs.~(\ref{r-e}), which define the singular points,
are:
\begin{equation} \label{K_E}
\rho_{{{\cal E}}} = l_{\phi}/{\cal K}_{{{\cal E}}},\,\,
{{\cal E}} = {{\varepsilon}} ({\cal K}_{\overline{{\cal E}}})
+ {\cal K}_{{{\cal E}}}/l_{\phi} \
\equiv  {\cal W}({\cal K}_{{{\cal E}}})\,,
\end{equation}
where ${\cal W}({\cal K})$ and ${\cal K}_{{{\cal E}}}$
are the dimensionless analogs of $W({\cal P})$ and ${\cal P}_E$
defined by Eqs.~(\ref{r-e}).
Bellow, ${{\cal E}}$ presents the dimensionless total energy
of the pair.
This formal scaling to the Bohr units will allow us to compare the results
with those known from the quantum Coulomb problem.

As a first example, we consider the simplest model of the MH type energy dispersion,
for which the dimensionless kinetic energy of relative motion of the electron pair is
\begin{equation}  \label{M-H}
{{\cal E}} (k) = - \frac{1}{2} k^2 + \beta k^4,\,\,\beta > 0\,.
\end{equation}
The corresponding characteristic parameters (see Fig.~\ref{fig-1} (b)) -are:
$k_i=1/2\sqrt{3 \beta},\,k_m=1/2 \sqrt{\beta},\,
k_0=1/\sqrt{2 \beta},\,{{\cal E}} (k_m) = 1/16 \beta$.
According to Eq.~(\ref{p-phi-c}), the critical angular momentum necessary
for the existence of the singular points and the closed trajectories equals
$l_{\phi} = l_{\phi,c} = 3 \sqrt{3 \beta}$. For this $l_{\phi}$, the singular point
in the $\{\rho,\,k_{\rho}\}$ phase-plane
arises at $k_{\rho,s}=0,\,\rho_{s,c}=2 \sqrt{3 \beta}$
(${{{\cal E}}}_{s,c} = {\cal W}(k_i) =1/48 \beta$).

If we assume $\beta =0.25$,
then we obtain: $k_i=0.58,\,k_m=1,\,k_0=
1.41,\,{{\cal E}}(k_m) = - 0.25$, $l_{\phi,c} = 2.6$,
the energy corresponding to the onset of the singular points is
${{{\cal E}}}_{s,c}= 0.08$, and the radius of the emerging circulate orbit
is $\rho_{s,c}\approx 7.6$. Setting $l_{\phi} =3$, we find the closed trajectories occur
in a finite energy interval,
$0.05 < {{{\cal E}}} <0.06$. In  real space, all closed trajectories are located in the
ring $4 \leq \rho \leq 7.5$.
Setting $\beta = 0.5$, we find $k_i=0.41,\,k_m=0.7,\,{{\cal E}}(k_m) =- 0.125$,
$l_{\phi,c} = 3.67,\,{{{\cal E}}}_{s.c}= 0.04$ and $\rho_{s.c} = 9 $.
Assuming $l_{\phi} =4$, we find for the closed trajectories
$0.031 \leq {{{\cal E}}} \leq 0.034$
and $8 \leq \rho \leq 13$. At $l_\phi=5$, for the same characteristics we find
$0 \leq {{{\cal E}}} \leq 0.02$ and $8.8 \leq \rho \leq 22$ for the closed
trajectories.

These estimates illustrate that, increasing $\beta$ and shortening
the negative effective mass portion of ${{\cal E}} (k)$ lead to larger critical
values of the angular momentum, $l_{\phi,c}$, greater
sizes of the coupled electron pair and lower its energies of  coupling.

As a second example of the energy dispersion of the MH type, we consider the lowest electron
(or hole) band of bigraphene subjected to a voltage applied across the graphene layers.
For this case, the one-particle energy is~[\onlinecite{bigraphene-1}]:
\begin{equation} \label{bi-gr-1}
\epsilon ({\bm p}) = \sqrt {g^{2}/2 + {V}^{2}/4+{p^2 v_F}^{2}-\sqrt {g^{4}/4+{p^2 v_F}^{2} \left( g^{2} + {V}^{2} \right) }}\,,
\end{equation}
here, $v_F$ is the Fermi-velocity parameter of the graphene, $g (\approx 0.4\,eV)$
characterizes interaction between
graphene layers, and $V$ is the voltage bias  applied across the layers.
Assuming bigraphene on a substrate with
a dielectric constant $\kappa_0$, for the Coulomb potential (\ref{coulomb}) we
obtain $\alpha = 2 e_0^2/(1+\kappa_0)$ with $e_0$ being the elementary charge.
Next, we introduce the scaling parameters as in Eqs.~(\ref{scaling})
with $M = g^{2}/4  v_F^{2} V$:
\begin{eqnarray} \nonumber
p_B=\frac{ e_0^2 g^2}{2 \hbar v_F^2 V (1+ \kappa_0)}\,, \,
 r_B = \frac{2 \hbar^2 v_F^2 V (1+ \kappa_0)}{ e_0^2 g^2 }\,,\,\\
 E_B = \frac{e_0^4 g^2}{ \hbar^2 v_F^2 V (1+ \kappa_0)^2}\,.\label{scaling-B-G}
 \end{eqnarray}
The dimensionless two-particle energy ${{\cal E}} (k) = 2 [\epsilon (k p_B)
- \epsilon (0)]/E_B$ at small $k$ behaves as  ${{\cal E}} (k) \approx - k^2/2 + ...$.

For further estimates, we set $V = 0.25\,eV$
 (corresponding energy gap of bigraphene is about 0.21\,eV) and $\kappa_0=3.9$,
the latter is valid for $SiO_2$ substrate.   We find
  \begin{eqnarray} \nonumber
 &M = 0.028\,m_0,\,\,E_B =  0.128\,eV,\, \\
 &p_B/\hbar = 2.18 \times 10^6\,cm^{-1} ,\, r_B = 4.58 \times 10^{-7}\,cm\,. \label{BG-estimates}
 \end{eqnarray}
Here $m_0$ is the free electron mass.
  The characteristic parameters of the two-particle kinetic
  energy ${{\varepsilon}} (k)$ are:
$ k_i=0.62,\,  k_m= 1.14,\, k_0= 1.74, {\varepsilon} (k_m) = -0.3$
(i.e.,\, $\approx 0.038\,eV$).
Then, we obtain the critical value of the angular momentum,
$l_{\phi,c} =2.5$, the energy corresponding to the onset of the singular points,
${{{\cal E}}}_{s,c}= {\cal W}(k_i)=0.09$,  and the radius of the emerging circulate
orbit, $\rho_s \approx 4.25$. Setting $l_{\phi} =3$,
we find that the closed trajectories occur in a finite energy interval,
$0.04 < {{\cal E}} <0.06$ at $0.16 < {\cal K}_{{{\cal E}}} < 0.83$ (see Eq.~(\ref{K_E})).
In  real space, all closed trajectories are located in the ring
$3.6 \leq \rho \leq 19$. Setting $l_{\phi} = 4$,
we found that such trajectories occur at $0 < {{{\cal E}}} <0.03$,
$ {\cal K}_{{{\cal E}}}< 0.6$ and the inter-particle distance $\rho > 6.6$.

These estimates show that semiclassical trajectories corresponding
to spatially bounded relative motion of the two electrons exist only for
finite  values of the angular momentum $l_{\phi}$. That is, {\it to be coupled the pair of electrons
has to rotate}. The energy of the coupled electrons is always positive and less than
$|{\varepsilon} (k_m)|$. Note, for any  energy corresponding to
a trajectory of coupled motion there
always exists  a trajectory of uncoupled motion. The trajectories
of coupled and uncoupled motion are well separated in the
$\{\rho,\,k_{\rho}\}$ phase-space.

\section{Quantum analysis} \label{Quantum analysis}

Foregoing semiclassical analysis allowed us to understand qualitative features of
relative motion of the electron pair with the MH energy dispersion. However,
as can be seen from numerical estimates show,
this analysis is not always adequate for quantitative conclusions.
Below we develop a quantum approach to the problem under consideration.

For the quantum analysis, we assume that spin-orbit interaction is negligible and
orbital motion and spin motion can be separated. First, we will focus on the orbital motion.

\subsection{Quantum equations for orbital motion of the electron pair}

We will use the dimensionless variables of Eq.~(\ref{scaling})
and, particularly, the Hamiltonian ${\cal H}_0$ with ${\cal E} (\hat{\bf k})$ being the
function of the momentum operator $\hat{{\bf k}}= - i \frac{\partial}{\partial {\bm \rho}}$.
Thus, in the coordinate representation  the Schr{\"o}dinger equation for relative motion
of the electron pair is~[\onlinecite{comment-3}]
\begin{equation} \label{Shr-1}
{\hat {\cal H}_0} \Psi({\bm \rho}) =\left[{\cal E} ({\hat {\bf k}}) + \frac{1}{\rho}\right] \Psi({\bm \rho}) = {\cal E} \Psi({\bm \rho})\,.
\end{equation}
Because of the circular symmetry of the problem, we will use the polar coordinates $\rho, \phi$,
then $\Psi ({\bm \rho}) = \Psi(\rho, \phi)$. The angular momentum operator is
$\hat{l} = - i \frac{\partial}{\partial \phi}$; its eigenfunctions
and eigenvalues are
\begin{equation} \label{l-f}
\Phi_l (\phi)= \frac{1}{\sqrt{2 \pi}} e^{-i l \phi},\,\,l=0, \pm 1, \pm 2\,,
\end{equation}
The operator $\hat{l}$ commutes with ${\cal H}_0$, thus we can set
\begin{equation} \label{orbital-psi}
\Psi(\rho, \phi) = R_l (\rho) \Phi_l(\phi)\,.
\end{equation}
Keeping in mind that the semiclassical analysis proves  the existence of
rotating coupled electron pairs, below we consider solutions to Eq.~(\ref{Shr-1}) with
$|l| \geq 1$.

To obtain equation for the radial functions, $R_l (\rho)$, we use the following relationship valid
for a two-dimensional system:
\begin{eqnarray} \nonumber
{\hat {\bm k}}^2 R_l (\rho) \Phi_l(\phi) \equiv - \Delta_2 R_l (\rho) \Phi_l(\phi) = \\
\Phi_l(\phi)  \, \left[ - \frac{1}{\rho}\frac{d}{d \rho} \rho \frac{d}{d \rho} + \frac{l^2}{\rho^2}\right] R_l (\rho)\,,\label{k2-operator}
\end{eqnarray}
where $\Delta_2$ is the two-dimensional Laplacian. Now, from Eq.~(\ref{Shr-1})
we can formally write down the {\it operator} equation for $R_l (\rho)$
\begin{equation} \label{Shr-2}
\left[{\cal E} \left(\sqrt{\left[ - \frac{1}{\rho}\frac{d}{d \rho} \rho \frac{d}{d \rho} + \frac{l^2}{\rho^2}\right]} \right) +
\frac{1}{\rho}\right] R_l(\rho) = {\cal E}_l R_l (\rho)\,,
\end{equation}
where generally dependent on $l$ eigenvalue, ${\cal E}_l$, is introduced.

To solve the operator equation (\ref{Shr-2}) we will use the eigenfunctions of
the operator $
\left[ - \frac{1}{\rho}\frac{d}{d \rho} \rho \frac{d}{d \rho} + \frac{l^2}{\rho^2}\right]\,,
$
which are known to be the Bessel functions, $J_l(q \rho)$,~[\onlinecite{Bessel-f-1}]:
\begin{equation}  \label{Bessel-1}
\left[ - \frac{1}{\rho}\frac{d}{d \rho} \rho \frac{d}{d \rho} + \frac{l^2}{\rho^2}\right] J_l (q \rho) = q^2 J_l(q \rho)\,.
\end{equation}
Note, for arbitrary operator of the kinetic energy ${\cal E} (\hat{\bf k})$,
 the radial function
of {\em free particles}  with an energy ${\cal E}$ also can be expressed via
the Bessel function:
\begin{equation} \label{free-particles}
{\cal R}_{l,{\cal E}} (\rho) = J_l (q_{\cal E} \rho)\,,\,\,\,
\end{equation}
where $q_{\cal E} $ solves the equation ${\cal E} (q_{\cal E}) = \cal E$.

Below we will use {\it Fourier-Hankel  transformation}~[\onlinecite{Bessel-f-2}]:
\begin{equation} \label{B_F}
R_l (q) = \int_0^{\infty} d \rho \rho R_l (\rho) J_l (q \rho)\,.
\end{equation}
It is important that, for functions in the form of Eqs.~(\ref{l-f}) and
(\ref{orbital-psi}) with integer $l$, standard two-dimensional
{\it Fourier transformation} between real space and
momentum space,
$$\Psi({\bm \rho}) = \frac{1}{2 \pi} \int d^2 q \Phi ({\bm q})\,,\,\,\,
\Phi ({\bm q}) = \frac{1}{2 \pi} \int d^2 \rho \Psi({\bm{ \rho}})\,,  $$
aslo leads to the relation (\ref{B_F}) for the radial functions
${\cal R}_l (\rho)$ and ${\cal R}_l (q)$. In particular, this is valid for
solutions of well known two-dimensional Coulomb problem with
attractive potential, $\overline{\cal R}_l (\rho)$ and
$\overline{\cal R}_l (q)$ (see, for example, Ref.~[\onlinecite{Shibuya-Portnoi}]).

Now we transform Eq.~(\ref{Shr-2}) to the integral equation
\begin{equation} \label{Shr-int-1}
\left[{\cal E} (q) - {\cal E}_l \right] R_l (q) + \int_0^{\infty} d q' Q_l (q,\,q') R_l (q') =0\,,
\end{equation}
with the symmetrical kernel
\begin{equation} \label{kernel-1}
Q_l (q',q) = \int_0^{\infty} d \rho J_l (q' \rho)  J_l (q \rho)\,.
\end{equation}
For $q' >q$, the kernel explicit form is~[\onlinecite{Bessel-f-1}]:
\begin{eqnarray} \nonumber
Q_l (q',q) = {\frac { \Gamma  \left( l+\frac{1}{2} \right) }{\sqrt {
\pi } \left( \Gamma  \left( 1+l \right)  \right)}}  {(q')}^{-1-l}{q}^{l}\, \times \\
_2F_1 \left( \left[\frac{1}{2},l+\frac{1}{2} \right], \left[1+l \right],{\frac {{q}^
{2}}{{(q')}^{2}}} \right)\,.  \label{kernel-2}
\end{eqnarray}
Here $\Gamma (x)$ and $_2F_1 \left([a,b],c, x \right)$ are the gamma and hypergeometric
functions, respectively.
For $q' < q$, the kernel can be obtained from Eq.~(\ref{kernel-2}) by permutation $q' \leftrightarrow q$.
This permutation recovers the symmetry property of the kernel $Q_l(q', q)$.

Eq.~(\ref{Shr-int-1}) is the equation for the radial function in momentum space.
It is important that in Eq.~(\ref{Shr-int-1}) ${\cal E} (q)$ is the usual function, which
according to scaling (\ref{scaling}) at small $q$ behaves as
${\cal E} (q) \approx - \frac{1}{2} q^2 + ...\,.$

\subsection{Approximate solutions for the wavefunctions and energies}
According to the above semiclassical analysis, in the same energy interval, where
closed trajectories (in the phase-plane) of the electron pair occur, there  also
exist trajectories of the uncoupled motion. Thus, in the quantum analysis, we can
expect {\it a quasi-resonant character}
of coupled states of the electron pair.  We will seek for solutions to
Eq.~(\ref{Shr-int-1}) in the form
\begin{equation} \label{sol-1}
R_l (q) = \sum_n B_n^{{l}} \overline{R}_{l,n} (q) + {\cal B}_l \,,
\end{equation}
where $\{\overline{R}_{l,n} (q) \}$ is a set of given functions describing localized
states. ${\cal B}_l $ is a small contribution of unbound (free) states  of the pair,
it will be analyzed in next subsection.

As a set of functions $\{ \overline{R}_{l,n} \}$, we select the solutions of the
two-dimensional Coulomb problem with the attractive potential.
For the momentum representation, detailed
analysis of this problem have been done in papers~[\onlinecite{Shibuya-Portnoi}].
Corresponding functions are solutions for the equation
\begin{equation} \label{Shr-int-2}
\left[\frac{1}{2}q^2 - \overline{{\cal E}}_{n} \right] \overline{R}_{l,n} (q) -
\int_0^{\infty} d q' Q_l (q,\,q') \overline{R}_{l,n} (q') =0\,,
\end{equation}
with $\overline{{\cal E}}_{n} = - 1/2(n+1/2)^2$
valid  at $n \geq |l|$, i.e.,  a given $n^{th}$-state of the
two-dimensional Coulomb problem is $(2 n + 1)$-fold degenerate.
The explicit expressions of normalized functions $\overline{R}_{n,l} (q)$ are
\begin{eqnarray} \nonumber
\overline{R}_{l,n} (q) = (-1)^l i^n 2 \sqrt{(2n +1) \frac{(n-|l|)!}{(n+|l|)!}}\, \times \\
\frac{q_n^2}{(q_n^2+q^2)^{3/2}}
{\cal P}^l_n \left[ \frac{q^2_n- q^2}{q_n^2+q^2} \right]\,,  \label{coulomb-f}
\end{eqnarray}
where ${\cal P}^l_n [x]$ is associated Legendre polynomials, $q_n = 1/(n+1/2)$.
 One can show that, in  momentum space these wavefunctions
are essentially concentrated at a range of small momenta, $q \leq 1/(n+1/2)$.
The degree of localization of the functions  in small $q$-range increases for larger
$l$ and $n$.

Return to solutions of Eq.~(\ref{Shr-int-1}) in form (\ref{sol-1}). Multiplying
Eq.~(\ref{Shr-int-2})
by $-1$ and comparing it with Eq.~(\ref{Shr-int-1}) we can see
that for ${\cal E}_l > 0$ both equations are quite similar, the only difference is more
complex dependence of ${\cal E} (q)$. Accordingly, for the coefficients $B_n^{{l}}$ in series (\ref{sol-1}),
 we obtain the following equations:
\begin{equation} \label{eq-for-Bn}
\left[ {\cal E}_l - |\overline{{\cal E}}_{l,n}|  \right] B_n^{{l}} - \sum_{n' \geq l} \Bigl \langle \overline{R}^{*}_{l,n} \left| {\cal E}(q) +
\frac{1}{2} q^2 \right| \overline{R}_{l,n'} \Bigr \rangle B_{n'}^{l} =0\,,\,
\end{equation}
where $n >l$ and  $\Bigl \langle \overline{R}^{*}_{l,n} \left|...\right| \overline{R}_{l,n'} \Bigr \rangle$ means the matrix element
calculated on the Coulomb states $\{l,n\}$ and $\{l,n'\}$.
In this formulation, difference of the eigenvalues of the considered
problem from the Coulomb ones is expressed via deviation of ${\cal E} (q)$ from
the parabolic law.

Eqs.~(\ref{eq-for-Bn}) can be solved numerically for a given ${\cal E} (q)$ dependence.
Below we present results obtained for ${\cal E} (q)$ corresponding to Eq.~(\ref{bi-gr-1})
at parameters discussed in Section \ref{semiclassical} (bigraphene on
$SiO_2$ substrate). The procedure of solution of Eq.~(\ref{eq-for-Bn}) is the following
(Galerkin) mothod.
We use a finite number of the functions,
$\{\overline{R}_{l,l},\overline{R}_{l,l+1},..\overline{R}_{l,l+\nu-1}\}$.  This gives us  $\nu$ approximate
eigenvalues designated as ${\cal E}_{l,N}^{(\nu)},\,N=l, l+1,l+2,...l+\nu-1$ and
corresponding eigenvectors $\{B_{N}^{l,(\nu)}\}$. Then, we increase the integer $\nu$ until we reach
the descripancy $\left|{\cal E}_{l,N}^{(\nu)} -{\cal E}_{l,N}^{(\nu+1)}\right|/{\cal E}_{l,N}^{(\nu)} < 0.001$,
for four  lower states with $N=l,..l+3$.
The obtained values ${\cal E}_{l,N}$ are presented in Table 1.
There, for comparison we also show four eigenvalues, $|{\cal E}_{N,Coul}|$,
of the two-dimensional Coulomb problem of Eq.~(\ref{Shr-int-2}).

Similarly to the Coulomb problem, one can consider $N$ as the "main quantum number".
However, unlike the Coulomb problem we obtain an {\it inverse series} of the energy
levels: all quasi-bound states are excited states and larger main  quantum number $N$
corresponds to smaller energy $E_{l,N}$. From the results of Table 1, one may see
that for the problem under consideration the $(2 N +1)$-fold angular momentum
degeneracy, that is characteristic for the Coulomb
problem,  is lifted, though two-fold degeneracy
of the states with $\pm l$ remains (at $l \neq 0$). For any $l$ we obtain
${\cal E}_{l,N} >|{\cal E}_{N,Coul}|$. This is due to the fact that the function
$|{\cal E} (q)|$ is always less than the kinetic energy for  the Coulomb case, $q^2/2$.
When $l,\,N$ increase, the found energies ${\cal E}_{l,N}$ become closer to the Coulomb
analogs. This can be understood since with increase in $l,\,N$,
the wavevectors actual for formation of the
quasi-coupled states become smaller and difference between functions $|{\cal E} (q)|$
and $q^2/2$ diminishes.

For given $l,N$, the obtained coefficients $\{B_{N}^{l}\}$ allow
us to calculate the radial wavefunctions, $R_{l,N} (q) $, and the total wavefunctions
in  momentum space
\begin{equation} {\label{psi-k-space}}
\Phi_{l,N} ({\bf q}) = \frac{1}{\sqrt{2 \pi}} e^{-i l \phi_q} R_{l,N} (q)\,,
\end{equation}
with $q$ and $\phi_q$ being the polar coordinates of the vector $\bf q$.

\begin{figure}
\includegraphics{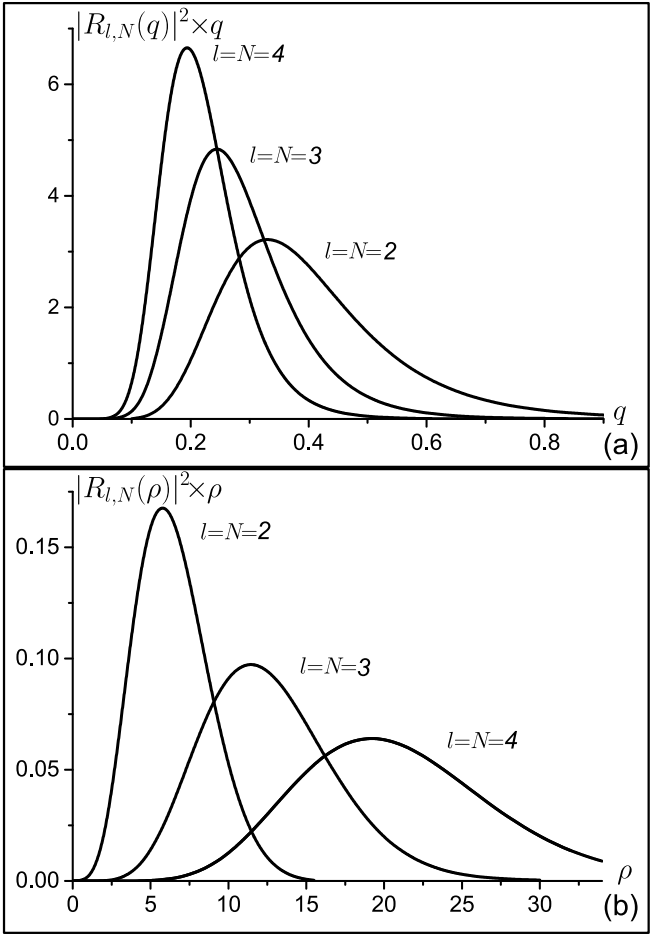}
\caption{{(a): Probability densities to find two electrons with the relative momentum $q$.
(b): Probability densities to find two electrons  at the relative distance $\rho$.
Results are shown for long living $\{l,\,N\}$ bi-electron states.}}
\label{fig-4}
\end{figure}

Probability to find the electron pair in $\{l,N\}$-bi-electron state with the
radial wavevector from the interval $q,\,q+dq$ is $|R_{l,N} (q)|^2 q dq$.
Corresponding probability density is presented in Fig.~\ref{fig-4}~(a) for several
long-living bi-electron states (see below the estimates of the decay time
of these states). One can see, that indeed
the bi-electron wavefunctions are concentrated at the region of
small $q$, where the effective mass of the
single-electron spectrum is essentially negative.
The larger quantum numbers $\{l,N \}$, the smaller actual $q$. We found average
values of $q$ for different $l, N$-states: $\overline{q}_{1,1}=0.65,\,
\overline{q}_{2,2}=0.3,\, \overline{q}_{3,3}=0.29,\,\overline{q}_{4,4}=0.22$.

The inverse Fourier-Hankel transformation gives the radial
wavefunctions in real space, $R_{l,N} (\rho)$. We used  these functions to
calculate the probability  to find two electrons in the interval
$\rho,\,\rho+d \rho$, i.e.,  $|R_{l,N} (\rho)|^2\rho d \rho$. Examples of corresponding
probability densities are shown in Fig.~\ref{fig-4} (b).
It is seen, that the probability densities are extended over large relative distances, $\rho$.
The average distances between the electrons in different $\{l,N\}$-states are:
$\overline{\rho}_{1,1}= 2.7,\,\overline{\rho}_{2,2}= 6.5,\,\overline{\rho}_{3,3}= 12.7,
\,\overline{\rho}_{4,4}= 22.5$, etc.

\subsection{Decay time calculations}
As stressed above, the found $\{ l,N\} $-bi-electron states have to be quasi-stationary. They decay to free particle
states of the same energy.  A formal approach to decay time calculations can be found
elsewhere~[\onlinecite{Vasko}].

According to this approach, to determine decay  of the found states we shall correct their
energies and wavefunctions. For example, in the coordinate representation we may use
corrected wavefunctions in the form
\begin{equation}
\Psi_{l,N} (\rho) = A_0 R_{l,N} (\rho) + \int d {\cal E} A_{\cal E} {\cal R}_{l,{\cal E}} (\rho)\,,
\end{equation}
 where ${\cal R}_{l,{\cal E}} (\rho)$ is the radial functions of free particles
 given by Eq.~(\ref{free-particles}),
 $A_0,\,A_{\cal E}$ are unknown coefficient and
 function. Then, applying the perturbation method
 to initial Eq.~(\ref{Shr-2})
 we find  these coefficients and an imaginary correction to the energy, $\gamma_{l,N}$:
\begin{equation} \label{gam}
{\cal E} \approx {\cal E}_{l,N}+ i \gamma_{l,N}\,,\,\, \gamma_{l,N} =  \pi  \frac{q_{l,N} |M_{l,N}(q_{l,N})|^2}{|d {\cal E}(q)/d q|_{q_{l,N}}}\,,
\end{equation}
where $q_{l,N}$ is defined by relationship ${\cal E} (q_{l,N}) = {\cal E}_{l,N}$ and
the matrix element is defined as follows
\begin{eqnarray} \nonumber
M_{l,N}(q)  =  \int_0^{\infty} d \rho J_l (q_n \rho) R_{l,N} (\rho)= \\
    \int_0^{\infty} d q' q' R_{l,N} (q') Q_l (q',q) \,.\nonumber
\end{eqnarray}
Calculated values of $\gamma_{l,N}$ are presented In Table 1.

Decay factor $\gamma_{l,N}$ determines broadening and the life time of the
two-electron coupled states,
$$\tau^{dec}_{l,N} =\frac{ \hbar}{2\gamma_{l,N}  E_B }\,.$$
Results obtained for $\gamma_{l,N}$ are valid if $\gamma_{l,N} \ll {\cal E}_{l,N}$,
when the energy level ${\cal E}_{l,N}$ is well defined.  Note, for $l=N=0$ we obtained
$\gamma_{l,N} \gtrsim {\cal E}_{l,N}$, which can be interpreted as the {\em nonexistence}
of such a state.
For $l \geq 1$, we obtain $\gamma_{l,N}/  {\cal E}_{l,N} < 1$. This ratio rapidly decreases
with increase in $l$. For $l,N \geq 3$, we find that $\gamma_{l,N}/  {\cal E}_{l,N} < 10^{-4}$.
This correlates with conclusions of the semiclassical analysis made
in Subsection \ref{N-estimates} for the same parameters:  semiclassical closed trajectories
arise at $l \geq 3$. In addition, for finite values of the angular momenta calculated
energies, ${\cal E}_{l,N}$, belong to the same energy intervals, for which  the
closed trajectories in the phase-plane are predicted.

\begin{table*}
Table~1:  Calculated energies, ${\cal E}_{l,N}$, broadenings, $\gamma_{l,N}$, and
total spins, $\Sigma$, of different quasi-coupled
states for the example of bigraphene on $SiO_2$ substrate discussed in the text.
Energies  of  the Coulomb states, $|{\cal E}_{N,Coul}|$, are presented for comparison.  \\
\vskip 0.2 cm
\begin{tabular}{|c|c||c|c|c||c|c|c||c|c|c||c|c|c||}    \hline \hline
 \,N \,& $|{\cal E}_{N,Coul}|$  &
 $ {\cal E}_{1,N}$ & $\gamma_{1,N} $ & $\Sigma$  &
$ {\cal E}_{2,N}$ & $\gamma_{2,N} $ & $\Sigma$  &
 $ {\cal E}_{3,N}$ & $\gamma_{3,N} $ & $\Sigma$ &
$ {\cal E}_{4,N}$ & $\gamma_{4,N} $ & $\Sigma$ \\  \hline \hline
1    &  0.222 &\, 0.337\, &\, 0.176\, & \,1 \, &       &       &     & & & & & & \\
2    & 0.08   & 0.102 & 0.028 & 1  & 0.094 & 0.001 & 0   &  & & & & & \\
3    & 0.041  & 0.048 & 0.007 &1   & 0.046 &
$4\,10^{-4}$ &0 & 0.044& $5\,10^{-6}$& 1 & & &\\
4   &0.025& 0.027 &0.002& 1 & 0.027& \,$1.5\,10^{-4}$\, &\,0\,&\, 0.026\,
&\,$3\,10^{-6}$\,&\,1\, &\, 0.026\,&\, $< 10^{-6 }$\,&\, 0\, \\
\hline \hline
\end{tabular}
\end{table*}

\subsection{Spin states of coupled electrons}

Consider briefly spin states of the coupled electron pair. We neglect
interaction between spins and orbital motion, that allows us to present
the total wavefunction of two electrons as a product of orbital and spin wavefunctions:
$${\bf \Psi} ({\bf r_1},{\sigma_1},{\bf r_2},{\sigma_2}) =
\Psi ({\bf r_1},{\bf r_2}) S({\sigma_1},{\sigma_2})\,,$$
where $\sigma_1, \sigma_2$ are the electron spins and $S({\sigma_1},{\sigma_2})$
is a spin function. The Pauli exclusion principle states that for two identical fermions
 the total wave function  is anti-symmetric with respect to
exchange of the particles.  It is easy to find symmetry properties of the orbital
wavefunction of Eq.~(\ref{orbital-psi}) with respect to permutation
${\bf r}_1 \rightarrow {\bf r}_2$. Indeed, such a permutation corresponds to
changing  ${\bf r} \rightarrow - {\bf r}$ or, in the polar coordinates, $\{\rho, \phi\}$,
to substitution $\phi \rightarrow \phi+\pi$.
The latter means that for the states with even angular momentum $l$ the
orbital wavefunction is symmetric, while those are antisymmetric for odd $l$.
Thus, to satisfy the Pauli exclusion principle for even $l$ the spin function
should be antisymmetric, which corresponds to total spin $\Sigma$ of the electron pair
equals 0. For odd $l$, the spin function should be symmetric, that implies
the total spin of the pair equals 1.

\subsection{Moving bi-electron}  \label{moving-bi}

So far, we considered that the center-of-mass of the electron pair is motionless, which is valid
when the total momentum of the pair, $\bf P$, is zero. If ${\bf P} \neq 0$, the electron pair is
moving  as a whole. For nonparabolic energy dispersion, $\epsilon ({\bf p})$, this translational motion and
relative motion of the electrons in the pair can not be separated.
However, assuming that $\bf P$ is small,  we can estimate the kinetic energy of the pair
and its effective mass. For that, we use the expansion of the Hamiltonian (\ref{H1}) in series
with respect to $\bf  P$:
\begin{eqnarray}\nonumber
H &\approx H_0 ({\bf p}, {\bf r}) + \frac{1}{4 p} \frac{d \epsilon}{d p} {\bf P}^2 +
\frac{1}{4 p^2}
\left[\frac{d^2 \epsilon}{d p^2} - \frac{1}{p} \frac{d \epsilon}{d p}  \right]
({\bf P p})^2 \equiv \\
& H_0 ({\bf p}, {\bf r}) + \delta H({\bf p},\,{\bf P})\,,\label{P-expansion-2}
\end{eqnarray}
where ${d \epsilon}/{d p}$ and ${d^2 \epsilon}/{d p^2}$ are functions of magnitude of the relative
momentum $\bf p$; $ \delta H({\bf p},\,{\bf P})$ is a correction to the Hamiltonian $H_0$
due to motion of the center-of-mass.
Using the scaling of Eqs.~(\ref{scaling}) we rewrite this correction as
$$
\delta {\cal H} ({\bf k}, {\bf Q})= \frac{\delta H}{E_B} =
\frac{1}{8 k} \frac{d {\cal E}}{d k} {\bf Q}^2 + \frac{1}{8 k^2}
\left[\frac{d^2 {\cal E}}{d k^2} - \frac{1}{k} \frac{d {\cal E}}{d k}  \right] ({\bf Q k})^2\,,
 $$
 with ${\bf Q} = {\bf P}/p_B$. Applying the perturbation theory
 method~[\onlinecite{Landau-QM}]
 for the $\{l,\,N\}$-state of the bi-electron, we obtain the energy correction in the
 form:
 \begin{eqnarray} \nonumber
 &\delta {\cal E}_{l,N} ({\bf Q}) = \frac{{\bf Q}^2}{2 {\cal M}_{l,N}}\,,\,\,\, \\
 &\frac{1}{{\cal M}_{l,N}} =
 \frac{1}{8} \int_0^{\infty} d q \left| R_{l,N} (q)  \right|^2 \left[\frac{d {\cal E}}{d q} + q \frac{d^2 {\cal E}}{d q^2} \right] \,, \label{kin-energy}
 \end{eqnarray}
where $\delta {\cal E}_{l,N} ({\bf Q})$ and ${\cal M}_{l,N}$ can be interpreted as
the dimensionless kinetic energy and the {\em effective mass} of the bi-electron
in the $\{l,N\}$-state.
Dependence of the effective mass of the bi-electron on its "internal" state is, obviously,
a manifestation of interaction of center-of-mass motion and internal relative motion.

Note, the wavefunctions $R_{l,N} (q)$ are mainly localized in the  range of small q,
where both terms in Eq.~(\ref{kin-energy}), $d {\cal E}/{d q} $ and $ {d^2 {\cal E}}/{d q^2} $,
are negative. This imply that the effective mass of the bi-electron should be also negative.

For the discussed example of bigraphene on the $SiO_2$ substrate,
the dimensionless effective masses of the bi-electron in different
$\{l,N\}$-states are:
${\cal M}_{1,1} =-16.6 ,\,{\cal M}_{2,2} = - 6,\,{\cal M}_{3,3} = -4.8,\,
{\cal M}_{4,4} = -  4.47$. Note, for scaling (\ref{scaling}), the dimensionless mass of
two noninteracting electrons with $q \rightarrow 0$ equals $-4$.
Thus bi-electrons are "heavy" particles with negative masses.

\section{Discussion} \label{Discussion}

In the present paper, we analyzed interaction of two electrons from the same
energy band with the complex single-particle energy dispersion.
Both semiclassical and quantum considerations showed that two electrons with the MH
single-electron energy dispersion, $\epsilon({\bf p})$, can form a composite
quasi-particle - the bi-electron. The coupling energies - energy of relative
motion of the two electrons - are positive (with respect to $2 {\epsilon (0) }$).
That is, the bi-electron corresponds to an excited state of the electron system.
Bi-electron coupled states exist in the continuum of the extended (free)
states of the electron pair. Thus, according to the quantum mechanics, the found
two-electron coupled states are of quasi-resonant character and have finite times of life.
Their times of life increase with the increase of angular momentum, thus more stable are
the rotating bi-electrons.
The semiclassical analysis also leads to stable coupled electron
pair only at finite values  of the angular momentum. Besides, the semiclassical analysis
showed that relative motion of two electrons is a peculiar character:
depending on the energy and the angular momentum (i.e., on the impact parameter)
e-e scattering can have three reversal
points and reversal points at non-zero radial momentum, etc.

The coupling energy, life-time, angular momentum, radius and other characteristics of the
bi-electrons are determined by the particular $\epsilon(p)$-dependence and dielectric
properties of the system. Let us discuss  these parameters for the above considered
example of bigraphene on the SiO$_2$ substrate in the transverse electric field.
Using the scaling  given in Eqs.~(\ref{scaling-B-G}), for relatively
stable bi-electron states one can find: $E_{2,2} = 12.4\,meV,\,\tau^{dec}_{2,2}=0.5\,ps,\,
$
$\overline{r}_{2,2}  \approx   29.8\,nm$; $E_{3,3} = 5.6\,meV,\,\tau^{dec}_{3,3}=1\,ns,\,
\overline{r}_{3,3}  \approx 58.2\,nm$; $E_{4,4} = 3.3\,meV,\,\tau^{dec}_{4,4} >5\,ns,\,
\overline{r}_{4,4} \approx 100\,nm$, where $E_{l,N},\,\tau^{dec}_{l,N}$ and
$ \overline{r}_{l,N}$ are the bi-electron energy, life-time and radius
in the $\{l,N\}$-state. As is seen, the quasi-coupled states with large angular momenta ($l \gtrsim 3$)
have macroscopically long life-times. Note, just these coupled bi-electron states are
allowed in the semiclassical theory. The sizes (radii) of these states are also large
($\overline{r} > 50\,nm$), they increase $\propto N$ at $N \gg 1$.

Note also, for the parameters discussed above,
the distances to the upper
valence band and the first excited electron band are estimated to be $0.21\,
eV$ and $0.31\,eV$, respectively, which
are much larger than the found coupling energies of the bi-electron,
$0.012...0.003\,eV$. That proves applicability of the single band approximation
used in this paper.

For the same bigraphene, varying the transverse electric field and/or dielectric
environment, one can change parameters of the bi-electrons. As an example, consider
briefly the free-standing bigraphene when one can set $\kappa_0 = 1$. Then, according to
the scaling (\ref{scaling-B-G})  at the voltage bias of the same amplitude
the coupling energies of the bi-electrons should be larger than those discussed above
by factor $(1+\kappa_{0,SiO_2})^2/4\approx 6$ and the corresponding radii of the
quantum states should be smaller by factor  $\approx 2.5$.

The bi-electron can move as a whole.  Such a translational motion can be
characterized by the total momentum of the composite particle.
At small value of the total momentum, the kinetic energy of the bi-electron is
a quadratic function of the momentum. Thus,  one can introduce
an effective mass of the bi-electron. Due to strongly nonparabolic
character of the energy dispersion, the translational motion is
coupled to relative motion of electrons composing the bi-electron.
As the result, the effective mass of the bi-electron depends
on its quantum state.   Using the scaling given by Eqs~(\ref{scaling})
and results of Subsection~\ref{moving-bi}, for the analyzed example of bigraphene
we obtain the following values of the effective mass: $-0.46\,m_0$ for $l=N=1$;
$-0.17\,m_0$ for $l=N=2$; $-0.13\,m_0$ for $l=N=3$ and $-0.12\,m_0$ for $l=N=4$, while the
effective mass of a single electron at $k=0$ equals $-0.056\,m_0$.
Thus the effective mass of the bi-electron is negative and considerably
varies with quantum numbers, $l, N$.

Above, properties of the bi-electron were illustrated for the example
of bigraphene in transverse electric field.
For this case, the MH single-electron energy spectrum is induced by
an external voltage. For realistic voltages, critical parameters of the MH spectrum
($p_m,\,\epsilon(p_m),\,M$) are such that energies
of coupled states are
relatively small ($\leq 10\,meV$) and radii of states are large ($\geq 50\,,nm$).
For two-dimensional crystals with {\it intrinsic} MH spectrum (for example, III-VI
compounds [\onlinecite{III-VI-compounds-3,III-VI-compounds-4}]),
the characteristic parameters are favorable for formation of the bi-particles
with larger coupling energies and stronger localization.

\section{Summary} \label{Summary}

We analyzed the interaction of a pair of the electrons, which  are characterized by the
mexican-hat single-electron energy dispersion. We have showed that relative motion
of  the electron pair is of a very peculiar character.  For example,
the real space trajectories corresponding to electron-electron scattering can have
three reversal points and reversal points at non-zero radial momentum.
These trajectories are strongly different from the usual ones.
Despite the repulsive Coulomb
interaction, two electrons can be coupled forming a composite particle - the bi-electron.
The bi-electron corresponds to excited states of two-electron system.
The bi-electron coupled states exist in continuum of extended (free)
states of the electron pair. Thus, the found bi-electron states are of quasi-resonant
character and have finite times of life.  We found that the rotating bi-electron is
the long-living composite particle. When spin-orbital interaction is negligibly
small, the bi-electron states with even angular momenta have zero spin (singlet states),
while those with odd angular momenta have unitary spin (triplet states).
The bi-electrons can be in translational motion. For slowly moving bi-electron,
 we have determined
the kinetic energy and the effective mass of the composite
particle.  Due to strongly nonparabolic
energy dispersion, the translational motion of the bi-electron is coupled to its
internal motion. This results in effective masses dependent on quantum states
of the bi-electron.

The studied rotating bi-electron replenishes the list of composite quasi-particles
with the Coulomb interaction which are already known for low-dimensional structures,
such as excitons, negatively and positively charged excitons
(trions)~[\onlinecite{Lampert,n-exciton,p-exciton}],
 and more complex fractionally-charged
quasi-particles (anyons) under the fractional Hall effect~[\onlinecite{FQHE}], etc.
Because there is a number of indications that the mexican-hat
single-electron energy dispersion occurs for novel two-dimensional materials,
we suggest that investigation of rotating bi-electrons may bring new interesting
effects in low-dimensional and low-temperature physics.


\begin{thebibliography}{100}
\bibitem{comment-2}
By simple substitutions $r \rightarrow r  p_{\phi},\,t \rightarrow t  p_{\phi}$ one can show that for
 a given energy dispersion $\epsilon(p)$,  the only actual parameter in Eqs~(\ref{r-t}), (\ref{pr-t}) and (\ref{phase-plane-1})
that has an effect on trajectories in the phase-plane, as well as in real space,
is $\alpha/p_{\phi}$.
\bibitem{comment-3}
The single energy band equation can be used for the analysis
of electron states with energies nearby this band, when
contributions of other remote energy bands are negligible.
\end{thebibliography}
\end{document}